\documentclass[superscriptaddress,secnumarabic,
amssymb,amsmath,nobibnotes,aps,prd,showkeys,showpacs,nofootinbib]{revtex4}%
\usepackage{graphicx}
\usepackage{float}
\usepackage{epsf}
\usepackage{bm}
\usepackage{amsmath}
\usepackage{amsfonts}
\usepackage{amssymb}
\usepackage{epstopdf}
\usepackage{natbib}
\usepackage{color}%
\providecommand{\U}[1]{\protect\rule{.1in}{.1in}}
\newcommand{\be}{\begin{equation}}
\newcommand{\ee}{\end{equation}}

\newcommand{\mincir}{\raise
-3.truept\hbox{\rlap{\hbox{$\sim$}}\raise4.truept\hbox{$<$}\ }}
\newcommand{\magcir}{\raise
-3.truept\hbox{\rlap{\hbox{$\sim$}}\raise4.truept\hbox{$>$}\ }}

\newtheorem{remark}{Remark}[section]

\begin{document}

\title{{ Reheating 
 formulas  in Quintessential Inflation via Gravitational Particle Production}}

\author{Jaume de Haro\footnote{E-mail: jaime.haro@upc.edu}}
\affiliation{Departament de Matem\`atiques, Universitat Polit\`ecnica de Catalunya, Diagonal 647, 08028 Barcelona, Spain}

\begin{abstract}
We calculate the reheating temperature in scenarios where heavy particles are gravitationally produced during a phase transition. We explore two distinct situations: the decay of these particles both during and after the kination phase. Subsequently, we determine the respective reheating temperatures. Finally, we constrain these temperatures based on considerations related to the overproduction of Gravitational Waves during the phase transition from the end of inflation to the onset of kination.

\end{abstract}

\vspace{0.5cm}


\maketitle

\thispagestyle{empty}

\section{Introduction}

The investigation into various reheating mechanisms within the framework of Quintessential Inflation has been a longstanding and intricate pursuit. Cosmologists have dedicated significant efforts to unravel the dynamics and implications associated with reheating processes specific to Quintessential Inflation.

\

Initially, the authors explore a mechanism rooted in the gravitational production of light particles, as elucidated in references \cite{pv, Spokoiny, Ford, Chung}. This mechanism, while providing a low reheating temperature around $10^3$GeV, encounters significant challenges, including the potential overproduction of Gravitational Waves (GWs) \cite{pv} and the effects of vacuum polarization during inflation \cite{fkl}. These challenges cast doubt on its viability. Consequently, an alternative mechanism, known as Instant Preheating \cite{fkl0, fkl, dimopoulos, campos}, emerges as a more promising solution. Grounded in the interaction between the inflaton field and a quantum scalar field, this approach circumvents the aforementioned issues.

\

Within the Quintessential Inflation framework, this alternative mechanism of reheating involves the creation of particles with an effective mass that grows over time. As these particles become non-relativistic, they undergo decay, contributing significantly to the reheating of the universe at an exceptionally high temperature, on the order of $10^9$ GeV \cite{Haro23} (Recall that in 
Super-Symmetric Theories, a reheating temperature below than $10^9$GeV is needed to avoid interferences with  the success of the Big Bang Nucleosynthesis (BBN), 
caused by the late decay of gravitationally interacting particles, such as the gravitino or
the moduli fields \cite{Ellis}. This is the so-called "gravitino problem"). This innovative perspective not only mitigates the challenges posed by the gravitational production of light particles but also introduces a more efficient reheating mechanism, leading to higher reheating temperatures. Only a potential challenge of this mechanism is that the coupling constant between the inflaton and the  quantum field 
responsible for particle production is highly constrained ranging  in a very narrow range \cite{Haro23}.

\

Concurrently, there has been a recent resurgence of interest in gravitational particle production, with a shift in focus towards the creation of heavy particles that subsequently decay into lighter ones, aligning with the Hot Big Bang model \cite{Hashiba, Ling, Lakinen, Haro}. Motivated by this perspective, the present work delves into this novel scenario, providing analytical formulas for the reheating temperature associated with the gravitational production of heavy particles. A comprehensive discussion ensues, evaluating temperatures around $10^7$ GeV, obtained when the created particles have heavy masses close to the scale of inflation,  which is close to  $10^{-6} M_{pl}$, and assessing their compliance with constraints arising from the potential overproduction of Gravitational Waves. This  exploration represents a significant advancement in understanding the  dynamics of the reheating mechanisms in Quintessential Inflation through the production of super-massive particles.

\

Throughout  the manuscript we  use natural units, i.e., 
 $\hbar=c=k_B=1$,  and the reduced Planck's mass is denoted by $M_{pl}\equiv \frac{1}{\sqrt{8\pi G}}\cong 2.44\times 10^{18}$ GeV.

\section{Reheating via gravitational production of heavy particles}

We begin by denoting $\rho_B(t)$ and $\langle\rho(t)\rangle$ as the energy densities of the background and the heavy massive particles produced gravitationally, respectively.

\

During the kination epoch, which in Quintessential Inflation takes place shortly after the end of inflation and  where all the energy of the background is kinetic, prior to the decay of the produced particles, using the sub-index $kin$ to signify the onset of kination, the evolution of the respective energy densities is:
\begin{eqnarray}
    \rho_B(t)=\rho_{B, kin}\left( \frac{a_{kin}}{a(t)}\right)^6, \qquad 
\langle \rho(t)\rangle= \langle \rho_{kin}\rangle   \left( \frac{a_{kin}}{a(t)}\right)^3,\end{eqnarray}
because
due to the significant mass, namely $m_{\chi}$,  of the particles under consideration, their energy density scales akin to non-relativistic matter. Specifically, during the kination phase, the background energy density scales proportionally to $a^{-6}$, while the energy density of the produced particles is calculated utilizing the $\beta$-Bogoliubov coefficients, as follows
\cite{Zeldovich}:
\begin{eqnarray}\label{vacuum-energy2}
\langle\rho(t)\rangle= \frac{1}{2\pi^2a^4(t)}\int_0^{\infty} k^2\omega_k(t)|\beta_k|^2 dk\cong \frac{m_{\chi}}{2\pi^2a^3(t)}\int_0^{\infty} k^2|\beta_k|^2 dk,
\end{eqnarray}
with $\omega_k(t)=\sqrt{k^2+m_{\chi}^2a^2(t)}$ being the frequency of the modes,  where we employ the approximation $\omega_k(t)\cong m_{\chi}a(t)$. The Bogoliubov coefficients are  subject to the following differential equations \cite{Zeldovich}:
\begin{eqnarray}\label{Bogoliubov}
\left\{ \begin{array}{ccc}
\dot{\alpha}_k(t) &=& \frac{\dot{\omega}_k(t)}{2\omega_k(t)}e^{-2i\int^{t} \frac{\omega_k(t)}{a(t) }dt}
\beta_k(t)\\
\dot{\beta}_k(t) &=& 
\frac{\dot{\omega}_k(t)}{2\omega_k(t)}e^{2i\int^{t}
\frac{\omega_k(t)}{a(t)}dt }
\alpha_k(t)\end{array}\right.
\end{eqnarray}
and the relationship $|\alpha_k(t)|^2-|\beta_k(t)|^2=1
$.  It is crucial to recognize that the $\beta$-Bogoliubov coefficients encapsulate both vacuum polarization effects and particle production. Shortly after the initiation of kination, the polarization effects become negligible. This implies that when they stabilize at a value denoted by $\beta_k$, the coefficients only reflect the contribution of the produced particles \cite{Haro21}.

\

At this juncture, it is important to note that solving (\ref{Bogoliubov}) requires an understanding of the evolution of the scale factor, which is derived from the semi-classical Friedmann equation: 
\begin{eqnarray}\label{Friedmann}
  \left(\frac{\dot{a}(t)}{a(t)}\right)^2\equiv  H^2(t)= \frac{1}{3M_{pl}^2}(\rho_B(t)+\langle\rho(t)\rangle) ,\end{eqnarray}
that incorporates the back-reaction stemming from particle production, indicating that the evolution of the scale factor is influenced by the Bogoliubov coefficients.

\

Fortunately, the approach to solving (\ref{Bogoliubov}) involves recognizing that well before the conclusion of kination, the energy density of the produced particles becomes subdominant. As a result, the semi-classical Friedmann equation can be approximated by the classical Friedmann equation $H^2(t)= \frac{\rho_B(t)}{3M_{pl}^2}$. Thus, the scale factor can be determined by solving the background equation:
\begin{eqnarray}
    \ddot{\phi}+\frac{\sqrt{3}}{M_{pl}}
    \sqrt{\frac{\dot{\phi}^2}{2}+V(\phi)}
    \dot{\phi}+V'(\phi)=0,
\end{eqnarray}
where the background is represented by the inflaton field $\phi$. Effectively,  once we have the evolution of the inflaton field up to the end of kination, one can calculate the energy density of the background. By inserting this value into the classical Friedmann equation, the evolution of the Hubble rate and, consequently, the scale factor can be determined well beyond the commencement of kination. This information is sufficient for numerical solutions of (\ref{Bogoliubov}) and the determination of the stabilized Bogoliubov coefficient.

\

However, numerically solving (\ref{Bogoliubov}) poses considerable difficulty, prompting the use of an analytical formula for the energy density of conformally coupled produced particles at the onset of kination \cite{haro23}
\begin{eqnarray}\label{rho1}
   \langle \rho_{kin}\rangle
   \cong
   \frac{1}{4\pi^3}
   e^{-\frac{\pi m_{\chi}}{2\sqrt{2}H_{END}}}
   \sqrt{\frac{m_{\chi}}{\sqrt{2}H_{END}}} H_{END}^2m_{\chi}^2, \end{eqnarray}
where $H_{END}$ denotes the value of the Hubble rate at the end of inflation, which can be calculated analytically by equating the slow-roll parameter 
$\epsilon=\frac{M_{pl}^2}{2}\left( \frac{V'(\phi)}{V(\phi)}\right)^2$ to $1$, what provides the value of the inflaton field at the end of inflation Additionally, one has to  
take into account that at the end of inflation one has:
\begin{eqnarray}
    \dot{\phi}_{END}^2=V(\phi_{END})\Longrightarrow
    H_{END}\cong \frac{1}{\sqrt{3}M_{pl}}\sqrt{\frac{\dot{\phi}_{END}^2}{2}+V(\phi_{END})}=\frac{1}{\sqrt{2}M_{pl}}\sqrt{V(\phi_{END})}.    \end{eqnarray}

\

After understanding the evolution of these energy densities before  the decay of the heavy particles into lighter ones, two distinct scenarios unfold: decay during and after the conclusion of the kination regime. Consequently, we will delve into both situations with  attention to detail.

\subsection{Decay before the end of kination}
Let $\Gamma$ be the decay rate of the heavy massive particles, and it is worth noting that the decay process concludes when $\Gamma$ is of the same order as the Hubble rate.

\

Denoting $\rho_{B, dec}$ and $\langle \rho_{dec}\rangle$ as the energy density of the background and that of the produced particles at the end of the decay, respectively, after the decay, they evolve as:
\begin{eqnarray}
    \rho_B(t)=\rho_{B, dec}\left( \frac{a_{dec}}{a(t)}\right)^6 \qquad \mbox{and} \qquad
\langle \rho(t)\rangle= \langle \rho_{dec}\rangle   \left( \frac{a_{dec}}{a(t)}\right)^4,\end{eqnarray}
because the background dominates the evolution during the kination and, after the decay,  the particles are currently  relativistic.

\

Hence, given the virtually instantaneous nature of the thermalization process, the universe undergoes reheating at the conclusion of kination, denoted by the sub-index $end$, that is, when both energy densities are of the same order. This leads to the relation:
\begin{eqnarray}
    \left( \frac{a_{dec}}{a_{end}}\right)^2
    =\frac{\langle \rho_{dec}\rangle}{\rho_{B, dec}} \Longrightarrow
\langle \rho_{end}\rangle= 
\frac{\langle \rho_{dec}\rangle^3}{\rho_{B, dec}^2}.  \end{eqnarray}

\

Therefore, from the Stefan-Boltzmann law, the reheating temperature has the following expression:  
\begin{eqnarray}\label{reheating0}
 T_{{reh}}\equiv \left(\frac{30}{\pi^2g_{reh}} \right)^{1/4}
 \langle\rho_{end}\rangle^{\frac{1}{4}}= 
 \left(\frac{30}{\pi^2g_{reh}} \right)^{1/4}
 \langle\rho_{dec}\rangle^{\frac{1}{4}}
 \sqrt{\frac{\langle\rho_{dec}\rangle}{\rho_{B,dec}}}, 
 \end{eqnarray}
 where $g_{reh}=106.75$ is the effective number of degrees of freedom for the Standard Model.

\

At this juncture, we can enhance this formula by considering the evolution of the corresponding energy densities before the decay. They follow the expressions:
\begin{eqnarray}
\rho_B(t)=\rho_{B,kin}\left( \frac{a_{kin}}{a(t)} \right)^6=
3H_{kin}^2M_{pl}^2 \left( \frac{a_{kin}}{a(t)} \right)^6,\qquad \langle \rho(t)\rangle=\langle \rho_{kin}\rangle \left( \frac{a_{kin}}{a(t)} \right)^3,
\end{eqnarray}
where  we have taken into account that $
\langle \rho_{kin}\rangle \ll \rho_{B,kin}$, implying
$\rho_{B,kin}\cong \langle \rho_{kin}\rangle + \rho_{B,kin}
=3H_{kin}^2M_{pl}^2$.

\

Then,  when the heavy particles have completely  decayed, which occurs when $H\sim \Gamma$,  the semi-classical Friedmann equation becomes: 
\begin{eqnarray}\label{semiclassical}
    3\Gamma^2M_{pl}^2=\rho_{B,kin}\left(\frac{a_{kin}}{a_{dec}}\right)^6+\langle\rho_{kin}\rangle
    \left(\frac{a_{kin}}{a_{dec}}\right)^3,\end{eqnarray}
which is  a quadratic equation in terms of $x=\left(\frac{a_{kin}}{a_{dec}}\right)^3$, and whose well-known solution is given by
\begin{eqnarray}\label{Theta11}
\left(\frac{a_{kin}}{a_{dec}}\right)^3=
\frac{1}{2H_{kin}}\left( \sqrt{H_{kin}^2\Theta^2+
{4\Gamma^2}}-H_{kin}\Theta  \right),
\end{eqnarray}
where  we 
have introduced  the concept of {\it heating efficiency } as \cite{Rubio}:
\begin{eqnarray}\label{Theta}
\Theta\equiv \frac{\langle \rho_{kin}\rangle}{\rho_{B,kin}}.
\end{eqnarray}

Using this definition,  we can write
\begin{eqnarray}\label{dec}
    \frac{\langle \rho_{dec}\rangle}{\rho_{B,dec}}=
    \frac{2H_{kin}\Theta}{\sqrt{H_{kin}^2\Theta^2+4\Gamma^2}-H_{kin}\Theta}\qquad 
    \mbox{and}\qquad
    \langle \rho_{dec} \rangle =\rho_{B,kin } \Theta 
    \frac{\sqrt{H_{kin}^2\Theta^2+4\Gamma^2}-H_{kin}\Theta}{2H_{kin}},\end{eqnarray}
and thus, 
taking into account that 
$\rho_{B,kin}\cong 3H_{kin}^2M_{pl}^2$, 
after some algebra,  the reheating temperature (\ref{reheating0}) is given by 
\begin{eqnarray}\label{reheating1}
    T_{reh}=\left(\frac{90}{\pi^2g_{reh}} \right)^{1/4}
\left(\frac{2H_{kin}\Theta}{\sqrt{H_{kin}^2\Theta^2+4\Gamma^2}-H_{kin}\Theta}  \right)^{1/4}\sqrt{\frac{\Theta H_{kin}}{M_{pl}}}M_{pl}.
\end{eqnarray}

\

On the other hand, given that the decay occurs during kination, we have the constraints $\Gamma \leq H_{kin}$ and $\langle \rho_{dec}\rangle\leq \rho_{B,dec}$, resulting in:
\begin{eqnarray}\label{constraint-gamma}
    \sqrt{2}\Theta H_{kin}
    \leq \Gamma\leq H_{kin}.
\end{eqnarray}

This is because, before the decay, the energy density of the produced particles scales as $a^{-3}$, leading to:
\begin{eqnarray}
    \langle \rho_{dec}\rangle\leq \rho_{B,dec}
    \Longrightarrow \langle \rho_{kin}\rangle\leq   
    \rho_{B,kin}\left(\frac{a_{kin}}{a_{dec}}\right)^3
    \Longrightarrow \Theta\leq\frac{1}{2H_{kin}}\left( \sqrt{H_{kin}^2\Theta^2+
{4\Gamma^2}}-H_{kin}\Theta  \right)\nonumber\\
 \Longrightarrow 3H_{kin}\Theta\leq\sqrt{H_{kin}^2\Theta^2+
{4\Gamma^2}}
\Longrightarrow 9H_{kin}^2\Theta^2\leq H_{kin}^2\Theta^2  +{4\Gamma^2}
\Longrightarrow 
2H^2_{kin}\Theta^2\leq \Gamma^2\Longrightarrow 
\sqrt{2}H_{kin}\Theta\leq \Gamma.\end{eqnarray}

Next, 
recognizing   that when the decay occurs well before to the end of kination  the decay rate satisfy $\Gamma\gg \sqrt{2}\Theta H_{kin}$, the reheating temperature will become: 
\begin{eqnarray}\label{reheating2}
    T_{reh}=\left(\frac{90}{\pi^2g_{reh}} \right)^{1/4}
\left(\frac{\Theta H_{kin}}{\Gamma}  \right)^{1/4}\sqrt{\frac{\Theta H_{kin}}{M_{pl}}}M_{pl}.
\end{eqnarray}

\

Additionally, the maximum reheating temperature is attained when $\Gamma= \sqrt{2}\Theta H_{kin}$ and the minimum one when $\Gamma= H_{kin}$, yielding the following expressions:
\begin{eqnarray}\label{Tmax}
    T_{reh}^{max}= 
    \left( \frac{90}{\pi^2 g_{reh}} \right)^{1/4}\sqrt{\frac{\Theta H_{kin}}{M_{pl}}}M_{pl}, 
    \qquad \mbox{and}\qquad
    T_{reh}^{min}=\left(\frac{90}{\pi^2g_{reh}} \right)^{1/4}
\left(\frac{2\Theta}{\sqrt{\Theta^2+4}-\Theta} 
\right)^{1/4}\sqrt{\frac{\Theta H_{kin}}{M_{pl}}}M_{pl},
\end{eqnarray}
where,  given that $\Theta\ll 1$, the minimum reheating temperature can be approximated as:
\begin{eqnarray}\label{Tmin}
    T_{reh}^{min}=\left(\frac{90 \Theta}{\pi^2g_{reh}} \right)^{1/4}\sqrt{\frac{\Theta H_{kin}}{M_{pl}}}M_{pl}.
\end{eqnarray}

\

Finally, also observe that to address the gravitino problem and ensure that the universe has been sufficiently reheated well before the BBN epoch, which occurs around $1$ MeV, these constraints must be imposed:
\begin{eqnarray}
    T_{reh}^{max}\leq 10^9\mbox{ GeV}\qquad \mbox{and}\qquad 
    T_{reh}^{min}\geq 1 \mbox{ MeV},\end{eqnarray}
what leads to the following initial bound for the heating efficiency
\begin{eqnarray}\label{bound_theta0}
    10^{-24}\leq \Theta\leq 10^{-12}.
\end{eqnarray}

\subsection{Decay after the end of kination}
In this subsection, we will examine the scenario where the decay occurs after the end of kination. In this case, given the instantaneous nature of the thermalization process, reheating is concluded upon the completion of decay. Therefore, the reheating temperature is determined by:
\begin{eqnarray} \label{reheating3}
T_{reh}=\left( \frac{30}{\pi^2 g_{reh}} \right)^{1/4}\langle\rho_{dec}\rangle^{1/4}, 
\end{eqnarray}
where one must ensure that ${\Gamma}\leq H_{end}$.  The value of the Hubble rate at the end of kination can be calculated  by considering that, in this scenario, the energy density of the produced particles decays as $a^{-3}$ throughout the entire kination phase. Thus, at the end of kination:
\begin{eqnarray}
\left( \frac{a_{kin}}{a_{end}} \right)^3=\frac{\langle\rho_{kin}\rangle}{\rho_{B,kin}}=\Theta\Longrightarrow 
H_{end}^2=\frac{2\rho_{B,end}}{3M_{pl}^2}= 
\frac{2\rho_{B,kin}}{3M_{pl}^2}\left( \frac{a_{kin}}{a_{end}} \right)^6=2H_{kin}^2\Theta^2,
\end{eqnarray}
and thus, the constraint $\Gamma\leq H_{end}$, obviously becomes
$\Gamma\leq \sqrt{2}\Theta H_{kin}.$

\

To refine the formula for the reheating temperature (\ref{reheating3}), we perform the following calculation:
\begin{eqnarray}\label{calculation}
\langle\rho_{end}\rangle=\rho_{B,end}=3H_{kin}^2\Theta^2M_{pl}^2 \Longrightarrow
     \langle\rho_{dec}\rangle^{1/4}=
     \langle\rho_{end}\rangle^{1/4}\left(\frac{a_{end}}{a_{dec}}\right)^{3/4} =3^{1/4}
     \sqrt{\Theta H_{in}M_{pl}}\left(\frac{a_{end}}{a_{dec}}\right)^{3/4}     .
\end{eqnarray}


\

Next, note that after kination, the potential of the inflaton field can be disregarded until the onset of matter domination, at which point it begins to be significant and eventually becomes dominant, leading the universe into the dark energy epoch. For this reason, the dynamical equation of the background
\begin{eqnarray}
    \dot{\rho}_B+3H(1+w_B)\rho_B=0\Longrightarrow
    d\rho_B+3(1+w_B)\rho_B da=0 \Longrightarrow
d\rho_B+6\rho_B da=0,
\end{eqnarray}
where the effective Equation of State parameter is
 $w_B=\frac{\dot{\phi}^2-2V(\phi)}{\dot{\phi}^2-2V(\phi)}\cong 1$, has the solution
\begin{eqnarray}
     \rho_B(t)=\rho_{B,end}\left(\frac{a_{end}}{a(t)}\right)^6     \end{eqnarray}

Therefore, when the decay is immediately  finished, 
introducing the notation $x=\left(\frac{a_{end}}{a_{dec}}\right)^3$,
the semi-classical Friedmann  is given by:
\begin{eqnarray}
    3\Gamma^2M_{pl}^2=\rho_{B,end}(x+x^2)\Longrightarrow
    x^2+x-\left(\frac{\Gamma}{\Theta H_{kin}}\right)^2=0
    ,
\end{eqnarray}
having as a solution
\begin{eqnarray}\label{Theta12}
\left(\frac{a_{end}}{a_{dec}}\right)^3=
\frac{1}{2H_{kin}\Theta}\left( \sqrt{H_{kin}^2\Theta^2+
{4\Gamma^2}}-H_{kin}\Theta  \right),
\end{eqnarray}
and thus, using the formulas (\ref{Theta12}) and (\ref{calculation}), 
the reheating temperature takes the following form:
\begin{eqnarray}\label{temperature_after_kination}
T_{reh}=    
    \left(\frac{90}{\pi^2g_{reh}} \right)^{1/4}
\left(\frac{\sqrt{H_{kin}^2\Theta^2+4\Gamma^2}-H_{kin}\Theta}{2H_{kin}\Theta}\right)^{1/4}\sqrt{\frac{\Theta H_{kin}}{M_{pl}}}M_{pl}.
\end{eqnarray}

\

\begin{remark}
    The formulas (\ref{reheating1}) and (\ref{temperature_after_kination}) match 
    when the decay is 
    at the end of kination, i.e., when $\Gamma=\sqrt{2}\Theta H_{kin}$,  obtaining the maximum reheating temperature
    (\ref{Tmax}).
\end{remark}

\

Finally, 
when the decay is well after the end of kination, i.e., when $\Gamma\ll \sqrt{2}\Theta H_{kin}$, the reheating temperature has the simple form:
\begin{eqnarray}\label{temperature_after_kination1}T_{reh}=    
    \left(\frac{90}{\pi^2g_{reh}} \right)^{1/4}\sqrt{\frac{\Gamma}{M_{pl}}}M_{pl},
\end{eqnarray}
and thus, for a decay well after the end of kination, 
 by enforcing $1 \mbox{ MeV}\leq T_{reh}$ to ensure reheating before the BBN, one obtains the bound:
\begin{eqnarray}\label{Gamma_bound}
  {6\times 10^{-43}M_{pl}\leq \Gamma
\ll \sqrt{2} \times 10^{-6} \Theta M_{pl}\Longrightarrow \Theta\gg 4\times 10^{-37}}.\end{eqnarray}

\subsection{Calculation of the heating efficiency}

Given that the reheating temperature is contingent on the value of $\Theta$, its calculation becomes imperative. To facilitate this computation, we employ the analytic formula for the energy density of conformally coupled particles produced at the initiation of kination (\ref{rho1}).

\

Assuming the conventional scenario where there is no significant drop in energy during the phase transition, i.e., $H_{END}\sim H_{kin}$ , we can derive:
\begin{eqnarray}\label{Theta1}
    \Theta
   \cong
   \frac{1}{12\pi^3}
   e^{-\frac{\pi m_{\chi}}{2\sqrt{2}H_{END}}}
   \sqrt{\frac{m_{\chi}}{\sqrt{2}H_{END}}} \left(\frac{m_{\chi}}{M_{pl}}\right)^2. \end{eqnarray}

We observe that its maximum value is attained when $m_{\chi}=\frac{5\sqrt{2}}{\pi}{H_{END}}$, yielding:
\begin{eqnarray}\label{Theta_max}
    \Theta_{max}=\frac{1}{6\pi^3}\left(\frac{5}{\pi e} \right)^{5/2}\left(\frac{H_{END}}{M_{pl}} \right)^2\cong 10^{-3}\left(\frac{H_{END}}{M_{pl}} \right)^2\cong 10^{-15},\end{eqnarray}
where we consider $H_{END}\sim 10^{-6}M_{pl}$. Consequently, considering (\ref{bound_theta0}) and (\ref{Theta_max}), we can deduce that when the decay occurs before the end of reheating, the heating efficiency falls within:
\begin{eqnarray}\label{Theta_range}
 10^{-24}\leq \Theta\leq 10^{-15},
\end{eqnarray}
and when the decay takes place well after the end of kination:
\begin{eqnarray}\label{Theta_range1}
 10^{-36}\ll \Theta\leq 10^{-15}.
\end{eqnarray}   
\section{Overproduction of Gravitational Waves}

This section is dedicated to presenting the constraints on Quintessential Inflationary models through the lens of Big Bang Nucleosynthesis. Here, we explicitly incorporate BBN constraints derived from the logarithmic spectrum of Gravitational Waves, and consequently, the BBN bounds arising from the potential overproduction of GWs.
\subsection{BBN bounds from the overproduction of GWs    }

The success of BBN imposes the condition \cite{pv}:
\begin{eqnarray}\label{bound_GW}
    \frac{\rho_{GW,reh}}{\langle\rho_{reh}\rangle}\leq {7\times 10^{-2}},
\end{eqnarray}
where $\rho_{GW}(t)$ represents the energy density of the GWs produced during the phase transition from the end of inflation to the onset of kination, and both quantities are assessed at the reheating time. The expression for the energy density of GWs is given by \cite{Ford}:
\begin{eqnarray}\label{GW}
\rho_{GW}(t)\cong 10^{-2}H_{kin}^4\left(\frac{a_{kin}}{a(t)}\right)^4.
\end{eqnarray}

\

\begin{remark}

Here, we can observe one of the issues with the gravitational production of light particles, as their energy density evolves according to \cite{pv}.
\begin{eqnarray}
    \langle \rho(t)\rangle\cong 10^{-2} N_s H_{kin}^4\left(\frac{a_{kin}}{a(t)}\right)^4,\end{eqnarray}
where $N_s$ is the number of produced scalar fields,  
 the bound (\ref{bound_GW}) becomes:
\begin{eqnarray}
    \frac{\rho_{GW,reh}}{\langle\rho_{reh}\rangle}=\frac{\rho_{GW,kin}}{\langle\rho_{kin}\rangle} = \frac{10^{-2}}{N_s} 
    \leq 10^{-2}\Longrightarrow \frac{1}{N_s}\leq 1.
\end{eqnarray}

However, for the minimal Gran Unification Theory  one has $N_s=4$, and thus, the bound is not surpassed. 
\end{remark}

\

Next, continuing with massive particles,  let us consider the constraints arising from the overproduction of GWs when the decay of the produced particles occurs during and after kination.

  \subsubsection{Decay during  kination}\label{before}
Since, after the decay, the energy density of the produced particles scales similarly to the energy density of GWs, we will have: 
\begin{eqnarray}
\frac{\rho_{GW,reh}}{\langle\rho_{reh}\rangle}=  \frac{\rho_{GW,dec}}{\langle\rho_{dec}\rangle}=
\frac{\rho_{GW,kin}}{\langle\rho_{kin}\rangle}
{\Theta}^{1/3}\left(\frac{ \sqrt{H_{kin}^2\Theta^2+
{4\Gamma^2}}- H_{kin}\Theta }{2H_{kin}\Theta}  \right)^{1/3}
\nonumber\\
\cong 3\times 10^{-3} \Theta^{-2/3}\left( 
\frac{ \sqrt{H_{kin}^2\Theta^2+
{4\Gamma^2}}- H_{kin}\Theta }{2H_{kin}\Theta}  \right)^{1/3}
\left( \frac{H_{kin}}{M_{pl}}\right)^2,
\end{eqnarray}
where we have utilized the formulas (\ref{Theta11}), (\ref{GW}), and the fact that at the onset of kination, the energy density of the produced particles is negligible compared to that of the background, i.e., $\rho_{B,kin}\cong 3H_{kin}^2M_{pl}^2$.

\

Now, since in the majority of models kination starts when $H_{kin}\sim 10^{-6}{M_{pl}}$, the bound (\ref{bound_GW})  becomes:
\begin{eqnarray}\label{bound_GW1}
    3\times 10^{-13} \Theta^{-2/3}
    \left( 
\frac{ \sqrt{H_{kin}^2\Theta^2+
{4\Gamma^2}}- H_{kin}\Theta }{2H_{kin}\Theta}  \right)^{1/3}
\leq { 7}.\end{eqnarray}

We will analyze this constraint when the reheating  temperature is close to its maximum and its minimum.

\begin{enumerate}
    \item 
{Decay at the end of kination kination.}

In this situation, the reheating temperature reaches its maximum, and $\Gamma\sim \sqrt{2}H_{kin}\Theta$. Then, the bound (\ref{bound_GW1}) becomes:
\begin{eqnarray}\label{bound_GW2}
    3\times 10^{-13} \Theta^{-2/3}\leq { 7}
    \Longrightarrow \Theta\geq {9\times 10^{-21}},
    \end{eqnarray}
which,  combined with the bound (\ref{Theta_range}),  leads to 
\begin{eqnarray}
   { 9\times 10^{-21}}\leq \Theta \leq 10^{-15}.
\end{eqnarray}

Applying this constraint to the maximum reheating temperature (\ref{Tmax}),  we obtain the following bound:
{
\begin{eqnarray}
    T_{reh}^{max}\geq 
    5 \times 10^{-14} M_{pl}\cong  10^5 \mbox{ GeV}.
    \end{eqnarray}}

In fact, the maximum value of this temperature is achieved when $\Theta=\Theta_{max}\Longrightarrow m_{\chi}=\frac{5\sqrt{2}}{\pi}H_{END}$, obtaining
\begin{eqnarray}\label{Tmax1}
    T_{reh}^{max}= 
    \left( \frac{90}{\pi^2 g_{reh}} \right)^{1/4}\sqrt{\frac{ H_{kin}}{10M_{pl}}} 10^{-7}M_{pl}\cong 2\times 10^{-11} M_{pl}\cong 5\times 10^7 \mbox{ GeV},
    \end{eqnarray}
    where, once again, we have taken $H_{kin}\sim 10^{-6} M_{pl}.$
Therefore, we have
\begin{eqnarray}
    {10^5} \mbox{ GeV}\leq T_{reh}^{max}\leq  5\times 10^7 \mbox{ GeV}. \end{eqnarray}
    
\item{Decay at the beginning of kination.}

In this  scenario,  the reheating temperature reaches its  minimum when
$\Gamma\sim H_{kin}$. Then,  since $\Theta\ll 1$, 
the bound (\ref{bound_GW1}) becomes:
\begin{eqnarray}\label{xXx}
    \Theta\geq 
   { 4\times 10^{-14}},
\end{eqnarray}
which is incompatible with the maximum value of the heating efficiency obtained in (\ref{Theta_max}). Therefore, we can conclude that the decay must be produced well before the onset of kination.

\end{enumerate}

\subsubsection{Decay after  kination}
First of all, we have to note that when the decay of the heavy particles is after the end of kination, the reheating concludes when the decay is completed. In addition, before the decay, the energy density of the produced particles scales as $a^{-3}$. Therefore, keeping this in mind, we can calculate
\begin{eqnarray}
\frac{\rho_{GW,reh}}{\langle\rho_{reh}\rangle}=  \frac{\rho_{GW,dec}}{\langle\rho_{dec}\rangle}= \frac{\rho_{GW,kin}}{\langle\rho_{kin}\rangle}\frac{a_{kin}}{a_{dec}}
=10^{-2}\frac{H_{kin}^4}{\Theta \rho_{B,kin}}\frac{a_{kin}}{a_{end}}
\frac{a_{end}}{a_{dec}}
=\frac{10^{-2}}{3}\Theta^{-1}
\frac{a_{kin}}{a_{end}}
\frac{a_{end}}{a_{dec}}\left( \frac{H_{kin}}{M_{pl}}\right)^2.\end{eqnarray}

Next, using, as we have already showed,  that when the decay is after kination
\begin{eqnarray}
    \left( \frac{a_{kin}}{a_{end}}\right)^3=\Theta\qquad \mbox{and}
   \qquad  \left( \frac{a_{end}}{a_{dec}}\right)^3=
   \frac{1}{2H_{kin}\Theta}\left( \sqrt{H_{kin}^2\Theta^2+
{4\Gamma^2}}-H_{kin}\Theta  \right)   
   \cong \left(\frac{\Gamma}{\Theta H_{kin}}\right)^2,
\end{eqnarray}
where, once again, we have assumed $\Gamma\ll \sqrt{2}\Theta H_{kin}$,
we obtain 
\begin{eqnarray}
\frac{\rho_{GW,reh}}{\langle\rho_{reh}\rangle}= \frac{10^{-2}}{3}\Theta^{-4/3}
\left({\frac{\Gamma}{H_{kin}}}\right)^{2/3}
\left( \frac{H_{kin}}{M_{pl}}\right)^2.\end{eqnarray}

And taking $H_{kin}\sim 10^{-6} M_{pl}$, the bound (\ref{bound_GW})
becomes
\begin{eqnarray}
{10^{-8}\left({\frac{\Gamma}{M_{pl}}}\right)^{2/3}
\leq 21\Theta^{4/3}\Longrightarrow
\Gamma\leq  10^{14}\Theta^{2}M_{pl}},
\end{eqnarray}
which combined with (\ref{Gamma_bound}) leads to 
\begin{eqnarray}\label{Theta13a}
  { 6\times 10^{-43}\leq 10^{14}\Theta^{2}\Longrightarrow
    \Theta\geq 10^{-28}},
    \end{eqnarray}
and thus,  when  the decay is after   kination, one has
\begin{eqnarray} \label{Theta13}
{  10^{-28}}\leq \Theta \leq 10^{-15}.
\end{eqnarray}

\subsection{BBN constraints from the logarithmic spectrum of GWs}

The logarithmic spectrum of GWs, namely $\Omega_{GW}$ defined as $\Omega_{GW}\equiv \frac{1}{\rho_c}\frac{d\rho_{GW}(k)}{d\ln k}$, where $\rho_{GW}(k)$ is the energy density spectrum of the produced GWs and $\rho_c=3H_0^2M_{pl}^2$ is the critical density. It is well-known that it scales as {$k$ during kination} \cite{Rubio}, producing a spike in the spectrum of GWs at high frequencies. Therefore, to prevent GWs from destabilizing the BBN, the following bound must be imposed \cite{Maggiore}:
\begin{eqnarray}\label{GW0}
    I\equiv h_0^2\int_{k_{BBN}}^{k_{END}}\Omega_{GW}(k) d\ln k\leq {5\times 10^{-6}}, 
\end{eqnarray}
where $h_0\cong 0.678$ parametrizes the experimental uncertainty in determining the current value of the Hubble constant, and $k_{BBN}$ and $k_{END}$ represent the momenta associated with the horizon scale at the BBN and the end of inflation, respectively. The primary contribution to the integral {(\ref{GW0})} arises from modes that exit the Hubble radius before the inflationary epoch and subsequently re-enter during the kination phase, i.e., for $k_{end}\leq k\leq k_{kin}$, where $k_{end}=a_{end}H_{end}$ and $k_{kin}=a_{kin}H_{kin}$. For these modes, one can calculate the logarithmic spectrum of GWs as outlined in \cite{Giovannini}
\begin{eqnarray}\label{GW1}
    \Omega_{GW}(k)=\tilde{\epsilon}\Omega_{\gamma}h^2_{GW}\frac{k}{k_{end}}
    \ln^2\left(\frac{k}{k_{kin}} \right), 
\end{eqnarray}
where $h^2_{GW}=\frac{1}{8\pi}\left( \frac{H_{kin}}{M_{pl}}\right)^2$ represents the amplitude of the GWs, $\Omega_{\gamma}\cong 3\times 10^{-5} h_0^{-2}$ is the present density fraction of radiation, and the quantity $\tilde{\epsilon}$—approximately equal to $0.05$ for the Standard Model of particle physics—accounts for the variation in massless degrees of freedom between decoupling and thermalization. Now, inserting expression (\ref{GW1}) into (\ref{GW0}) and neglecting the sub-leading logarithmic terms, one arrives at the bound
\begin{eqnarray}\label{constraint_log}
   { 2\times} 10^{-2}\left(\frac{H_{kin}}{M_{pl}}\right)^2\frac{k_{kin}}{k_{end}}\leq 1.
\end{eqnarray}

\

{

On the other hand, 
to compare the bounds (\ref{bound_GW}) and (\ref{GW0}), start by noting that $I=h_0^2\frac{\rho_{GW,0}}{\rho_c}$, with $\rho_{GW,0}$ representing the energy density of gravitational waves at the present time. Let $\langle \rho_0\rangle$ be the current energy density of matter. Then, express $I$ as follows:
\begin{eqnarray}
I=\frac{\rho_{GW,0}}{\langle \rho_0\rangle}h_0^2\Omega_{m,0},
\end{eqnarray}
where $\Omega_{m,0}=\frac{\langle \rho_0\rangle}{\rho_c}$ stands for the density parameter for matter.

\

Since the evolution of the energy densities of the produced particles and gravitational waves  is the same during radiation, we have $\frac{\rho_{GW,reh}}{\langle \rho_{reh}\rangle}= \frac{\rho_{GW,eq}}{\langle \rho_{eq}\rangle}$. Here, the sub-index $eq$ indicates quantities evaluated at matter-radiation equality. After matter-radiation equality, the energy density of produced particles scales as $a^{-3}$, yielding:

\begin{eqnarray}
\frac{\rho_{GW,0}}{\langle \rho_{0}\rangle}=
\frac{\rho_{GW,eq}}{\langle \rho_{eq}\rangle}\frac{1}{1+z_{eq}}=
\frac{\rho_{GW,reh}}{\langle \rho_{reh}\rangle}\frac{1}{1+z_{eq}},
\end{eqnarray}
where $z_{eq}=\frac{a_0}{a_{eq}}-1$ is the redshift at matter-radiation equality. Consequently, we obtain:

\begin{eqnarray}
I=\frac{\rho_{GW,reh}}{\langle \rho_{reh}\rangle}\frac{ h_0^2\Omega_{m,0} }{1+z_{eq}}\cong 4\times 10^{-5}
\frac{\rho_{GW,reh}}{\langle \rho_{reh}\rangle}
,
\end{eqnarray}
where we used $z_{eq} = 3387$ and $h_0^2\Omega_{m,0} = 0.1424$, as provided in the last column of Table 2 in \cite{Planck}. Upon inserting the bound (\ref{bound_GW}), we obtain $I \leq 3 \times 10^{-6}$, which is of the same order as the constraint (\ref{GW0}). This suggests that (\ref{bound_GW}) practically imposes equally restrictive conditions on the heating efficiency as (\ref{GW0}), as we will demonstrate in \ref{before}.

\

Finally, it is important to note that in \cite{Giovannini,Giovannini1}, to calculate (\ref{GW1}), the author considers a model with a transition from kination to radiation. This implies that the produced particles are either very light, or if they are heavy, they must decay during kination. If the decay occurs after kination, the transition would involve kination, a brief period of matter-domination, and then radiation. In other words, the bound (\ref{constraint_log}) only holds when the decay occurs during kination. As we will see in \ref{after}, the constraint (\ref{constraint_log}), applied to particles decaying after the end of kination, leads to a more restrictive bound on the heating efficiency than (\ref{bound_GW}), which contradicts our earlier explanation.

\

}



\subsubsection{Decay before the end of kination}\label{before}

First we start with the formula 
\begin{eqnarray}
    3H_{end}^2M_{pl}^2=2\rho_{B,end}\Longrightarrow H_{end}=\sqrt{2}H_{kin}\left(
    \frac{a_{kin}}{a_{end}}\right)^3,
\end{eqnarray}
to obtain
\begin{eqnarray}
    \frac{k_{kin}}{k_{end}}=
    \frac{a_{kin}H_{kin}}{a_{end}H_{end}}=    \frac{1}{\sqrt{2}}\left(
    \frac{a_{kin}}{a_{end}}\right)^{-2}=\frac{1}{\sqrt{2}}    \left(
    \frac{a_{dec}}{a_{end}}\right)^{-2}\left(
    \frac{a_{kin}}{a_{dec}}\right)^{-2}.
    \end{eqnarray}
Next,  we use that 
\begin{eqnarray}\label{log_x}
\left(
    \frac{a_{dec}}{a_{end}}\right)^{2}=\frac{\langle \rho_{dec}\rangle}{\rho_{B,dec}}=\frac{2H_{kin}\Theta}{\sqrt{H_{kin}^2\Theta^2+{4\Gamma^2} }-H_{kin}\Theta    },
\end{eqnarray}
where we have used the formula (\ref{dec})  and also  
(\ref{Theta}),  to get:
\begin{eqnarray}
\frac{k_{kin}}{k_{end}}= \frac{1}{\sqrt{2}} \Theta^{-2/3}\left( 
\frac{\sqrt{H_{kin}^2\Theta^2+{4\Gamma^2} }-H_{kin}\Theta    }
{2H_{kin}\Theta}\right)^{1/3}.
\end{eqnarray}

Inserting it into (\ref{constraint_log}), for $H_{kin}\sim 10^{-6} M_{pl}$, we obtain 
\begin{eqnarray}
{{\sqrt{2}}} \Theta^{-2/3}\left( 
\frac{\sqrt{H_{kin}^2\Theta^2+{4\Gamma^2} }-H_{kin}\Theta    }
{2H_{kin}\Theta}\right)^{1/3}\leq 10^{14}.
\end{eqnarray}

As we have already shown in \ref{before}, we need to assume that the decay is close to the end of kination. Choosing
$\Gamma\sim \sqrt{2}H_{kin}\Theta$, we find
\begin{eqnarray}
   { \Theta\geq 2\times 10^{-21}},
\end{eqnarray}
which is { is of the same order as}  (\ref{bound_GW2}).


\

{

\

Similarly, when the decay occurs near the beginning of kination, the bound (\ref{GW0}) results in $\Theta \geq 10^{-14}$, which is also of the same order as (\ref{xXx}).

\

Therefore, as we have already argued, the condition (\ref{bound_GW}) imposes the same constraint on the heating efficiency as (\ref{GW0}).

}

\subsubsection{Decay after the end of kination}\label{after}

In this case, during all the kination period the energy density of the produced particles scales as $a^{-3}$, what implies
\begin{eqnarray}
    \Theta=\left(\frac{a_{kin}}{a_{end}} \right)^3\qquad \mbox{and} \qquad
    H_{end}=\sqrt{2}H_{kin}\Theta.
\end{eqnarray}

Therefore, 
\begin{eqnarray}
    \frac{k_{kin}}{k_{end}}=
    \frac{a_{kin}H_{kin}}{a_{end}H_{end}}=    
    \frac{1}{\sqrt{2}}\Theta^{-2/3},
\end{eqnarray}
and  the bound (\ref{constraint_log}), will become:
\begin{eqnarray}
 {\sqrt{2}\times}  {10^{-2}}\left(\frac{H_{kin}}{M_{pl}} \right)^2 \Theta^{-2/3}\leq 1,
\end{eqnarray}
which for $H_{kin}\sim H_{END}\sim 10^{-6}{M_{pl} }$, leads to 
{
\begin{eqnarray}\label{Theta3b}
    \Theta\geq {2}\times 10^{-21},
\end{eqnarray}
}which is more restrictive than the bound (\ref{Theta13a}). {(
Recall that (\ref{Theta13a}) is obtained by imposing that the decay occurs well after the end of kination, i.e., by imposing $\Gamma\ll \sqrt{2}\Theta H_{\text{kin}}$. Additionally, $\Theta\sim 10^{-28}$ is obtained for a reheating temperature close to $1$ MeV.}

\

{

However, as we have previously explained, we cannot apply the bound (\ref{GW0}) together with (\ref{GW1}) when the decay is after kination. This is because the formula (\ref{GW1}) is derived under the assumption that radiation immediately follows kination. In cases where the decay of heavy particles occurs well after kination, there is an intermediate period of matter-domination between kination and radiation, making the application of (\ref{GW1}) inappropriate. In conclusion, the fact that the bound (\ref{Theta3b}) is more restrictive than (\ref{Theta13a}) does not imply that the constraint (\ref{GW0}) is more restrictive than (\ref{bound_GW}) when the decay occurs after kination. This discrepancy is due to the incorrect application of the formula (\ref{GW1}).

\

}


{
\subsection{Bounds coming from the overproduction}

In summary, considering the constraints arising from the overproduction of GWs, we can draw the following conclusions:}
\begin{enumerate}
    \item 
When the decay is close to the end of kination, the heating efficiency is bounded by    \begin{eqnarray}
       {9\times 10^{-21}}\leq \Theta\leq 10^{-15}, 
    \end{eqnarray}
which corresponds to masses in the range   
\begin{eqnarray}
    2.2\times 10^{-6}\leq \frac{m_{\chi}}{M_{pl}}\leq {1.9}\times 10^{-5},
\end{eqnarray}
leading to a maximum reheating temperature (\ref{Tmax}) within the bounds
\begin{eqnarray}
{10^5} \mbox{ GeV}\leq T_{reh}^{max}\leq  5\times 10^7  \mbox{ GeV},\end{eqnarray}
and a minimum temperature (\ref{Tmin}) satisfying 
\begin{eqnarray}
    10 \mbox{ GeV}\leq T_{reh}^{min}\leq  { 7}\times 10^3 \mbox{ GeV}.\end{eqnarray}

    \item When the decay is well after the end of kination, the 
    heating efficiency is bounded by 
    \begin{eqnarray}
       { 10^{-28}}\leq \Theta\leq 10^{-15},
    \end{eqnarray}\end{enumerate}
which is satisfied for masses in the range
\begin{eqnarray}
    2.2\times 10^{-6}\leq \frac{m_{\chi}}{M_{pl}}\leq {3.7}\times 10^{-5},
    \end{eqnarray}
and thus,  for a decay well after the end of kination, i.e., for  $\Gamma\ll \sqrt{2}\Theta_{max} H_{kin}$ and
taking into account 
the formula (\ref{temperature_after_kination1}),  the 
reheating temperature is bounded by 
\begin{eqnarray}
    1 \mbox{ MeV}\leq  T_{reh}\ll  5\times 10^7 \mbox{ GeV}.\end{eqnarray}

\section{Conclusions}

We have developed analytical formulas, specifically (\ref{reheating1}), (\ref{reheating2})-(\ref{Tmin}) and (\ref{temperature_after_kination}), for the reheating temperature within the framework of Quintessential Inflation, an scenario where the universe undergoes reheating post-inflation through the gravitational production of heavy particles conformally coupled with gravity. These heavy particles subsequently decay into lighter ones, contributing to the thermalization of the universe.  One of the aspects of our invertigation lies in considering the timing of this decay, whether it occurs during or after the kination epoch, leading to  distinct reheating temperatures.

\

These reheating formulas are intricately linked to   the concept of heating efficiency, defined as the ratio, at the initiation of kination, between the energy density of the produced particles and that of the background. Additionally, these formulas are influenced by the decay rate of the heavy particles and the Hubble parameter at the onset of kination. 
To ensure the viability of the model, we have established
bounds on the heating efficiency, stemming from constraints aimed at mitigating the overproduction of Gravitational Waves.

\

Once these bounds are ascertained, they furnish the means to constrain not only the reheating temperature but also the masses of the produced particles. This comprehensive analysis unravels  the intricate interplay between various factors influencing the reheating process in Quintessential Inflation, specifically through  gravitational particle production.

\section*{Acknowledgments} 
This work  
is supported by the Spanish grant 
PID2021-123903NB-I00
funded by MCIN/AEI/10.13039/501100011033 and by ``ERDF A way of making Europe''.

\end{document}